\documentclass[twocolumn,showpacs,bibnotes]{revtex4}
\usepackage{graphicx,epsfig}
\usepackage{dcolumn}

\begin{document}

\title{Forcing and Velocity Correlations in a Vibrated
Granular Monolayer}

\author{Alexis Prevost}
\email[Email: ]{prevost@physics.georgetown.edu}
\author{David A. Egolf}
\altaffiliation{Also, Center for Nonlinear Studies, Los Alamos
National Laboratory, Los Alamos, NM 87545}
\author{Jeffrey S. Urbach}

\affiliation{Department of Physics, Georgetown University,
Washington D.C., 20057}
\date{\today}

\begin{abstract}
The role of forcing on the dynamics of a vertically shaken
granular monolayer is investigated. Using a flat plate, surprising
negative velocity correlations are measured. A mechanism for this
anti-correlation is proposed with support from both experimental
results and molecular dynamics simulations. Using a rough plate,
velocity correlations are positive, and the velocity distribution
evolves from a gaussian at very low densities to a broader
distribution at high densities. These results are interpreted as a
balance between stochastic forcing, interparticle collisions, and
friction with the plate.
\end{abstract}


\maketitle

Granular gases, systems of large numbers of macroscopic grains in
rapid motion and interacting through dissipative collisions,
appear in a wide range of industrial applications and natural
phenomena. Energy must be supplied externally to compensate for
the inelastic collisions, so granular gases are necessarily
systems out of equilibrium. As a result, they may display dramatic
non-equilibrium effects such as non-gaussian velocity
distributions
\cite{OlafsenUrbach,Losert99,Rouyer00,VanNoije98,ChrisBizon,NieBenNaimChen}
and long-range spatial velocity correlations
\cite{VanNoije99,Pagonabarraga01,MoonShattuckSwiftPRE2001,BlairKudrolliRPRE2001}.
Non-gaussian velocity distributions are a direct demonstration of
the inapplicability of the Gibbs distribution, and significant
velocity correlations indicate the absence of `molecular chaos',
which is a crucial approximation normally used to solve the
Boltzmann equation and to calculate other fundamental quantities
in kinetic theory. Recent theoretical work has focused on the
non-equilibrium steady state obtained when the energy supplied by
spatially homogeneous random external forcing is balanced by the
dissipation due to the collisions such that the average energy of
the system remains constant
\cite{VanNoije98,VanNoije99,Pagonabarraga01}. Non-gaussian
velocity distributions and algebraically decaying velocity
correlations arise as a direct consequence of the energy
injection. Non-gaussian velocity distributions and velocity
correlations have been observed in a number of experiments
\cite{OlafsenUrbach,Losert99,Rouyer00,BlairKudrolliRPRE2001}, but
in each case the forcing is sufficiently different from that of
the theoretical models that a direct comparison is difficult. In
this Letter, we provide a direct demonstration of the determining
role that the forcing plays on the spatial velocity correlations
in a homogeneously forced granular gas. We will also show how the
framework of kinetic theory can provide a coherent description of
the origin of the observed velocity distributions and
correlations.

We have investigated a quasi-2D granular system consisting of a
layer of a large number of spherical particles partially covering
a vertically driven horizontal plate. Two plates with different
surface properties were used in the experiments: a smooth circular
plate (20 cm in diameter) made of black anodized aluminum, and a
rough hexagonally shaped plate (30 cm between opposite corners).
The roughness of the latter is provided by a close-packed lattice
of blackened steel balls (1.19 mm diameter) glued to a flat plate.
The granular gas is made of uniform stainless steel spheres with
diameter $\sigma=1.59$\,mm (smooth plate), or $\sigma=3.97$ mm
(rough plate). On the smooth plate, the layer is constrained from
above by an anti-static coated plexiglas lid, 1.7 $\sigma$ above
the plate. Using an electromagnetic shaker, the plate is driven
sinusoidally, and the granular layer is brought to a
non-equilibrium steady state. The strength of the shaking $\Gamma
= A \omega^2/g$ is directly measured with a fast response
accelerometer ($A$ is the amplitude and $\omega$ the frequency of
the plate oscillation, and $g$ the acceleration due to gravity).

A high resolution camera (Pulnix TM1040) placed above the plate
records the bright spots on the tops of the balls produced by a
stroboscopic LED array. Instantaneous horizontal velocities ${\bf
v}_i$ are obtained from the displacements between strobe pulses.
The strobe is synchronized with the camera so that the first pulse
occurs at the end of the exposure of one frame, and the next pulse
occurs at the beginning of the subsequent frame. The time interval
between pulses ($\sim$ 1 ms) is chosen to be significantly smaller
than the mean collision time. The width of both pulses is
typically less than 0.05 ms. Displacements of the centers of the
bright spots are measured with 0.1 pixel accuracy. To eliminate
systematic errors in the velocity correlations, snapshots are
taken at a fixed phase relative to the driving signal
corresponding to the maximum average position of the layer.

The general behavior of the granular layer on the smooth plate has
been reported previously \cite{OlafsenUrbach} and a rich phase
diagram depending on both $\Gamma$ and $\omega$ was found. Here we
focus mainly on the fluidized regime, obtained for $\Gamma\geq 1$.
All results presented in this Letter were obtained with
$\Gamma=1.5$ and $\omega/2\pi=60$ Hz.

The longitudinal and transverse velocity correlations, $C_{||}$
and $C_{\bot}$, respectively,  are calculated by
\begin{eqnarray}\label{velocitycorrdef}
C_{||,\bot}(r)= \sum_{i\neq j}^{N_r} v_i^{||,\bot}
v_j^{||,\bot}/N_r, \nonumber
\end{eqnarray}
where the sum runs over the $N_r$ pairs of particles separated by
a distance $r$, $v_i^{||}$ is the projection of ${\bf
v}_i$ along the line connecting the centers of particles $i$ and
$j$, and $v_i^{\bot}$ is the projection perpendicular to that
line.

\begin{figure}
\centerline{\epsfig{file=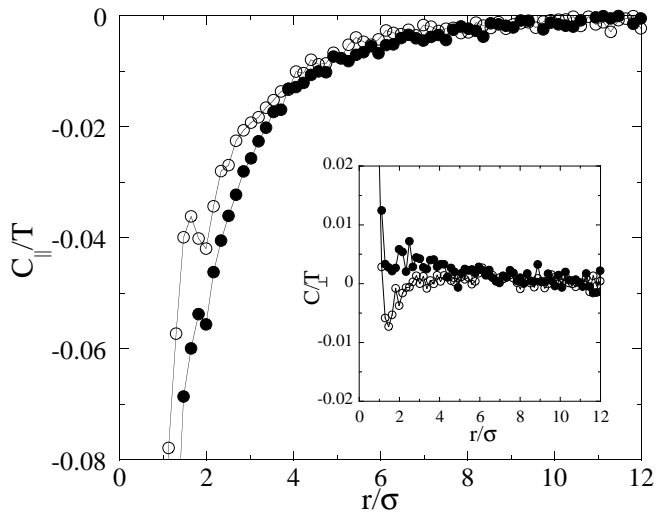,width=8.6cm}}
\caption{Longitudinal velocity correlations normalized by the
granular temperature $T$ on the smooth plate: $\Gamma=1.5$,
$\rho=0.4\,(\bullet)$ and $\rho=0.5\,(\circ)$. Inset: transverse
velocity correlations. Data represent averages over approximately
8000 pairs of images with about 180 balls per
image.}\label{fig:figure1}
\end{figure}

Figure \ref{fig:figure1} shows $C_{||}$ (inset: $C_{\bot}$)
normalized by the granular temperature $T=\langle{\bf
v}_i^2\rangle$ obtained on the smooth plate for $\rho=0.4$ and
0.5, and $\Gamma=1.5$ \cite{coverage}. Surprisingly, the
longitudinal velocity correlations are strongly negative for all
$r$ and all accessible densities: the velocities are
anti-correlated. $C_{||}$ decays slower than exponentially, and in
the range $r=[2\sigma,16\sigma]$ can be reasonably well described
by $C_{||}(r)\propto r^{-2}$. The structure visible between
$1\sigma$ and $2\sigma$ is stronger at high density, and its
origin is not understood. The transverse component $C_{\bot}$ is
much smaller and has a shorter range than $C_{||}$ (note the
difference in vertical scale). Nearly identical correlations were
observed without the lid in place.  We have performed molecular
dynamics simulations that closely match the conditions of the
experiment \cite{computer} and find qualitatively similar
correlations.

\begin{figure}
\centerline{\epsfig{file=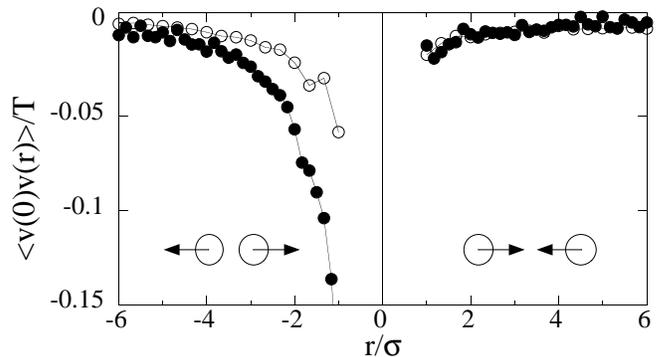,width=8.6cm}} \caption{
$\langle{\bf v}(0)\cdot{\bf v}(r)\rangle$ (as described in the
text) normalized by $T$ for $\Gamma=1.5$ and $\rho=0.4$
($\bullet$), $\rho=0.5$ ($\circ$).}\label{fig:figure2}
\end{figure}

The existence of long-range velocity correlations in a randomly
forced granular gas has been explained theoretically using a
hydrodynamic approach \cite{VanNoije99}.  The velocity
correlations result from the balance between the excitation of
hydrodynamic modes by the random external forcing and their
dissipation through diffusion. At shorter range, velocity
correlations arise from an increased probability of re-collision
due to the forcing \cite{Pagonabarraga01}.  This suggests that the
correlations could be quite sensitive to the particular mechanism
of energy injection, and we believe that this is the origin of the
dramatic difference between the negative correlations observed in
our experiment and the positive correlations observed in the
randomly forced model systems
\cite{VanNoije98,VanNoije99,Pagonabarraga01,MoonShattuckSwiftPRE2001}.
In those systems, energy is injected through white noise forcing
of each particle independently, while on the smooth plate energy
is injected only into the vertical motion of the spheres and is
then transferred to the horizontal motion through inter-particle
collisions. As a result of this transfer, collisions will result,
on average, in an increase in the magnitude of the relative
horizontal velocities ${\bf v}_r$ of pairs of particles. This
effect can be measured directly in the molecular dynamics
simulations, and at $\rho=0.5$ we find that the average value of
$\langle {\bf v}_r^2 \rangle$ for pairs of particles leaving a
collision is 13\% higher than for pairs entering a collision.
Experimental support for this scenario can be found in Fig.\
\ref{fig:figure2}, which shows a calculation of $\langle{\bf
v}(0)\cdot{\bf v}(r)\rangle$, where the average includes only
particles restricted to a narrow band along the direction of
motion of the particle at the origin, either ahead (for  positive
$r$) or behind (for negative $r$). There is a clear asymmetry with
respect to zero, showing that balls moving away from each other
contribute significantly more to the velocity anti-correlation
than do particles moving towards each other. The velocity
anti-correlations would presumably dissipate through diffusion as
the momentum gets transferred to surrounding particles through
collisions, thus it is not surprising that the observed
correlations extend over several mean free paths. This mechanism
for the generation of anti-correlated velocities may also operate
in inhomogeneously forced granular media when the average kinetic
energy of the grains is anisotropic.

To test our hypothesis about the origin of the correlations, we
have changed the forcing by using a rough surface, so that the
vibrating plate also injects energy directly into the horizontal
motion of the spheres. This energy input mechanism is closer to
the white noise forcing of Ref.\ \cite{VanNoije99}, and allows for
a more direct comparison between experiment and theory
\cite{chaotic}.

\begin{figure}
\centerline{\epsfig{file=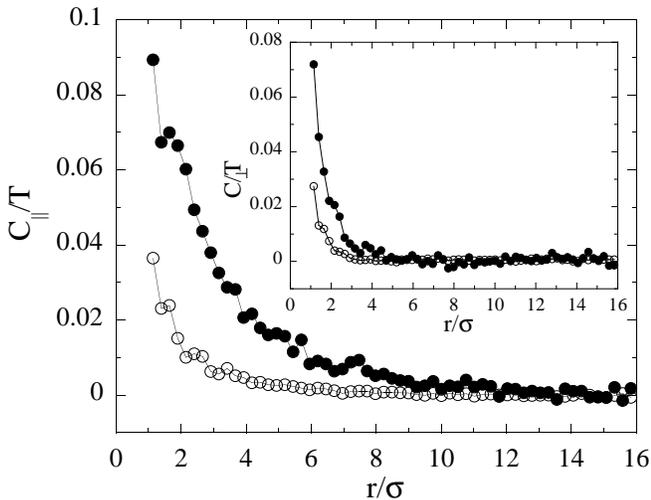,width=8.6cm}}\caption{Longitudinal
velocity correlations normalized by the granular temperature on
the rough plate: $\Gamma=1.5$, and $\rho=0.125\,(\bullet)$,
$\rho=0.6\,(\circ)$. Inset: transverse velocity correlations. Data
are averaged over 16000 pairs of images ($\sim$ 80 balls per
image) for $\rho=0.125$, and 8000 pairs ($\sim$ 420 balls) for
$\rho=0.6$. }\label{fig:figure3}
\end{figure}

The velocity correlations measured on the rough plate are shown on
Fig.\ \ref{fig:figure3} for $\Gamma=1.5$ \cite{Cluster}, and two
extremal densities. We find that both $C_{||}$ and $C_{\bot}$ are
positive for all measured $\rho$. $\langle{\bf v}(0)\cdot{\bf
v}(r)\rangle$, calculated as described above, is positive
everywhere and is roughly symmetric about zero. Molecular dynamics
simulations of the layer on the rough plate \cite{computer} show
similar correlations, and the effect of the collisions is to
decrease the average value of $\langle {\bf v}_r^2 \rangle$ by
3.7\% at $\rho = 0.5$. Thus in the absence of large vertical to
horizontal collisional energy transfer, the velocity
anti-correlations are replaced by positive velocity correlations,
presumably arising from the random forcing as described in Refs.
\cite{VanNoije99,Pagonabarraga01}. The dramatic difference in the
correlations between the smooth and rough plates is a clear
indication that the forcing, rather than the inelastic
interparticle collisions, determines the velocity correlations.

\begin{figure}
\centerline{\epsfig{file=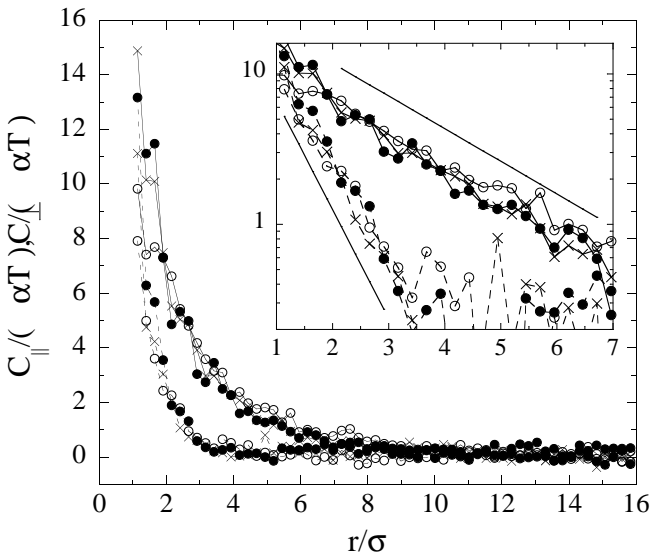,width=8.6cm}}\caption{Scaled
longitudinal and transverse velocity correlations normalized by
$T$. $C_{||}/(\alpha T)$ (symbols with solid lines) and
$C_{\bot}/(\alpha T)$ (symbols with dashed lines) for
$\Gamma=1.5$, $\rho=0.125$ ($\circ$), $\rho=0.5$ ($\times$), and
$\rho=0.6$ ($\bullet$). Inset: Log-linear plot of the scaled
velocity correlations (the straight lines are exponentials with
decay lengths of $2\sigma$ and $0.6\sigma$.)}\label{fig:figure4}
\end{figure}

Interestingly, we find that $C_{||}$ and $C_{\bot}$ can be roughly
scaled on two different curves of the form $C_{||,\bot} = T
\alpha(\rho) f_{||,\bot}(r)$ (Fig.\ \ref{fig:figure4}). The
functions $f_{||}$ and $f_{\bot}$ have different $r$ dependence,
but the function $\alpha(\rho)$ is the same for both $C_{||}$ and
$C_{\bot}$. $f_{||}$ and $f_{\bot}$ are reasonably well described
by $f(r) \propto e^{-r/r_o}$, with $r_o=2\sigma$ for $f_{||}$ and
$0.6\sigma$ for $f_{\bot}$. It is surprising that this decay
length has little density dependence, despite the fact that the
mean free path estimated from Enskog-Boltzmann kinetic theory
varies from 2.3$\sigma$ for $\rho=0.125$ to 0.16$\sigma$ for
$\rho=0.6$. A similar scaling relation is predicted in the
randomly forced model at large $r$ \cite{VanNoije99}, but the
scaling is the same for both $C_{||}$ and $C_{\bot}$, and $f(r)$
decays algebraically.

\begin{figure}
\centerline{\epsfig{file=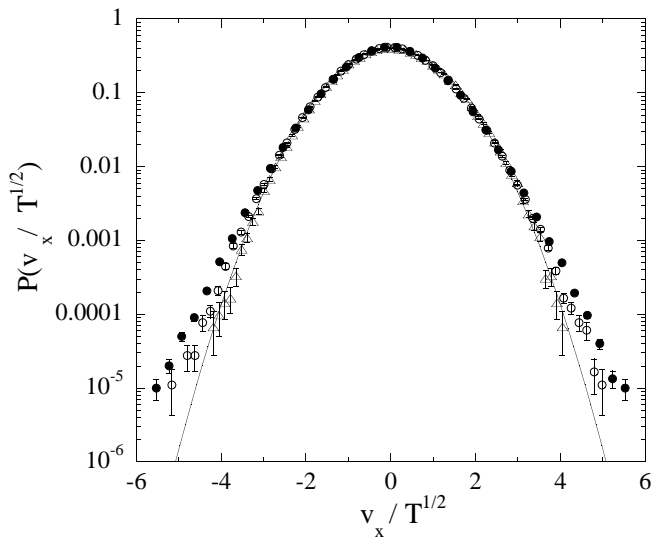,width=8.6cm}}\caption{Normalized
velocity distributions $P(v_x)$, on the rough plate for
$\Gamma=1.5$ and $\rho=0.014$\,$(\triangle)$,
$\rho=0.125\,(\circ)$, $\rho=0.6\,(\bullet)$. The solid line is a
gaussian. (Results for $P(v_y)$ are identical).
}\label{fig:figure5}
\end{figure}

An important difference between our experiment and the randomly
forced model is revealed by the velocity distributions $P(v)$ of
the balls on the rough plate. Figure \ref{fig:figure5} shows the
normalized $P(v_x)$ obtained for $\Gamma=1.5$, and $\rho$ ranging
from $0.014$ to $0.6$. At the lowest density, $P(v_x)$ is well
described by a gaussian. As $\rho$ increases, the tails of the
distribution rise significantly above a gaussian. At high density,
the velocity distribution resembles those reported previously on
flat plates \cite{OlafsenUrbach,Losert99}. To better understand
the effect of density on $P(v)$, we have followed the trajectories
of individual balls using a fast camera (838 frames/s,
\cite{OlafsenUrbach}), and measured the time-dependent velocity
autocorrelation function $C_{v}(t)=\langle{\bf v}(0)\cdot {\bf
v}(t)\rangle$ and find that $C_v(t)\approx e^{-t/ \tau(\rho)}$ at
all densities. The scattering length defined by
$l\equiv\sqrt{T}\tau(\rho)$ varies from about 0.5$\sigma$ at
$\rho=0.6$ to about 2.5$\sigma$ at $\rho=0.014$. At the lowest
density, $l$ is much smaller than the mean free path for ball-ball
scattering ($\sim$ 20 $\sigma$) obtained from kinetic theory. This
suggests that it is ball-plate scattering that is responsible for
the decay of the velocity auto-correlation function, and that the
scattering must therefore have a viscous component. The evolution
of $P(v)$ for a single ball on the rough plate can thus be very
likely modelled with an equation including only random white noise
forcing and a viscous term, such as the Fokker-Plank equation for
Brownian motion, which produces a gaussian $P(v)$. When the random
forcing is instead balanced by ball-ball scattering in a modified
Boltzmann equation, there is an over-population of the tails of
the distributions, $P(v) \sim \exp(-|v|^{3/2})$ \cite{VanNoije98}.
This is consistent with the experimentally measured $P(v)$ at high
densities, where the scattering is dominated by ball-ball
collisions. At intermediate densities both scattering processes
contribute and the evolution of $P(v)$ with density can likely be
described with a Boltzmann equation consisting of random white
noise forcing, collisional scattering, and a density independent
viscous drag. This model has been recently investigated
\cite{Puglisi99}, although primarily at higher inelasticities,
where significant clustering is observed.

The ball-plate scattering may also explain the lack of significant
density dependence of the decay length of the velocity
correlations. We speculate that the density independent scattering
with the plate, which unlike ball-ball collisions does not
conserve momentum, is more effective at destroying the long-range
velocity correlations, and therefore controls the observed decay.
Support for this scenario is provided by the fact that the decay
length for the exponentially decaying $C_{||}(r)$ at {\em all}
measured densities is close to the ball-plate scattering length
deduced from the individual trajectories at very low densities
(2$\sigma$ versus $2.5\sigma$). However, a full quantitative
understanding of the velocity correlations observed on the rough
plate will require a theoretical analysis along the lines of Ref.\
\cite{VanNoije99}, including the effect of viscous drag. Our
results provide a clear experimental demonstration that the
velocity distribution and correlations in a granular gas are not
determined exclusively by internal dynamics (inelastic ball-ball
collisions) but by a balance between internal dynamics and
external forcing.

The authors gratefully acknowledge Mickey Moore for providing us
with the rough plate. This work was supported by grants from the
Research Corporation, the Petroleum Research fund, DOE, and Grant
No. DMR-9875529 from the NSF. J.S.U. acknowledges support from the
Sloan foundation.


\begin{thebibliography}{14}
\bibitem{coverage} $\rho=N/N_{c}$, $N$ is the number of balls and
$N_{c}$ the number in a full close-packed monolayer.
\bibitem{computer} The interaction rules and parameters for the
molecular dynamics simulation were taken from Ref.
\cite{NieBenNaimChen}. The roughness was simulated by adding a
layer of close-packed spheres rigidly attached to the plate. The
values for $\sigma$, $\omega$ and $A$ are taken from the
experiment.
\bibitem{Cluster} At $\Gamma=1.5$, the layer is fully fluidized.
Clustering, due to trapping of the balls by the lattice, is
present for values of $\Gamma$ below 1.2.
\bibitem{chaotic} Although the roughness of the plate is periodic, its wavelength is considerably
less than the measured scattering length, and the motion of the
balls on the surface is presumably chaotic, so stochastic forcing
and scattering is likely to be a good approximation.
\end{thebibliography}
\end{document}